\documentclass[prl,english,twocolumn,amsmath,amssymb,superscriptaddress]{revtex4-1}
\usepackage{amsmath}
\usepackage{graphicx}
\usepackage{amssymb}
\usepackage{dsfont}
\usepackage{color}
\usepackage[bookmarks=true,colorlinks,linkcolor=blue,urlcolor=blue,citecolor=blue]{hyperref}

\usepackage{hyperref}
\usepackage{tikz}
\usetikzlibrary{calc}

\newcommand{\be}{\begin{equation}}
\newcommand{\ee}{\end{equation}}
\newcommand{\bea}{\begin{eqnarray}}
\newcommand{\eea}{\end{eqnarray}}

\newcommand{\mc}{\mathcal}

 \def       \im{\dot{\imath}}

\begin{document}

\title{Gapless excitations in non-Abelian  Kitaev spin liquids  with line defects}

\author{Lucas R. D. Freitas}
\affiliation{Departamento de F\'isica Te\'orica
e Experimental, Universidade Federal do Rio Grande do Norte, 
Natal, RN, 59078-970, Brazil}

\author{Rodrigo G. Pereira}
\affiliation{Departamento de F\'isica Te\'orica
e Experimental, Universidade Federal do Rio Grande do Norte, 
Natal, RN, 59078-970, Brazil}
\affiliation{International Institute of Physics, Universidade Federal do Rio Grande do Norte, 
Natal, RN, 59078-970, Brazil}

\begin{abstract}
We show that line defects in a non-Abelian Kitaev spin liquid harbor gapless one-dimensional Majorana modes if the  interaction across the defect falls  below a critical value. Treating the weak interaction  at the line defect within a mean-field approximation, we  determine the critical  interaction  strength  as a function of the external magnetic field. In the gapless regime, we use the low-energy effective field  theory to calculate the spin-lattice relaxation rate for a  nuclear spin near the defect and find a cubic temperature dependence that agrees with   experiments in the  Kitaev material $\alpha\text{-RuCl}_3$. 
\end{abstract}
\maketitle

\emph{Introduction.---}The Kitaev honeycomb model \cite{Kitaev2006} provides a prominent example of a quantum spin liquid  \cite{Savary2016,Knolle2019,Broholm2020} in which   spins  fractionalize into emergent Majorana fermions. In the presence of a magnetic field, the phase diagram of the model    includes a non-Abelian phase characterized by gapped bulk excitations and chiral edge states \cite{Kitaev2006}. The observation that an extended  Kitaev model can be realized in  strongly spin-orbit-coupled Mott insulators \cite{Jackeli2009,Chaloupka2010,Rau2014} led to the discovery of candidate materials, including  the iridates \cite{Singh2012,Chun2015,Kitagawa2018} and $\alpha$-RuCl$_3$ \cite{Plumb2014,Kim2015, Banerjee2016,Hentrich2020}. In the latter, the suppression of long-range zigzag order above  a critical value of an in-plane  magnetic field \cite{Banerjee2018} has been interpreted in terms  of a field-induced gapped spin liquid, with   supporting evidence from thermal Hall \cite{Kasahara2018,yokoi2020halfinteger} and specific heat measurements \cite{tanaka2020}.

The inevitable presence of defects in   real materials   both complicates and enriches the physics of Kitaev spin liquids \cite{Willans2010,Vojta2016,Knolledisorder,Yamada2020,Andrade2020,Kao2021,Nasu2021}.  Quite generally, disorder tends to modify   the low-energy density of states  in a way that  may overshadow  universal properties predicted 
 for the clean system.  For instance, single vacancies and magnetic impurities can  bind vortices of the   $\mathbb Z_2$ gauge field and  Majorana zero modes \cite{Willans2010,Vojta2016}.  A finite density of vacancies and bond randomness can account for the divergent low-energy density of states in  H$_3$LiIr$_2$O$_6$ \cite{Kitagawa2018,Knolledisorder}.

In this work, we investigate line defects, such as  dislocations and grain boundaries \cite{solyom-solids},  in the non-Abelian Kitaev spin liquid. Such one-dimensional (1D) defects can be engineered in monolayers of 2D materials \cite{Lin2016} and  their orientation depends on strain \cite{Komsa2013}. In bulk crystals, partial dislocations naturally appear bordering stacking faults \cite{solyom-solids}, which are ubiquitous in $\alpha$-RuCl$_3$  due to the weak van der Waals bonding between layers \cite{Johnson2015,Kim_Kee2016,Cao2016}.  Dislocations in the gapped Abelian phase of the anisotropic Kitaev model were studied in Refs. \cite{Petrova2013,Petrova2014}.

\begin{figure}[b]%
    \centering
       \includegraphics[width=.95\columnwidth]{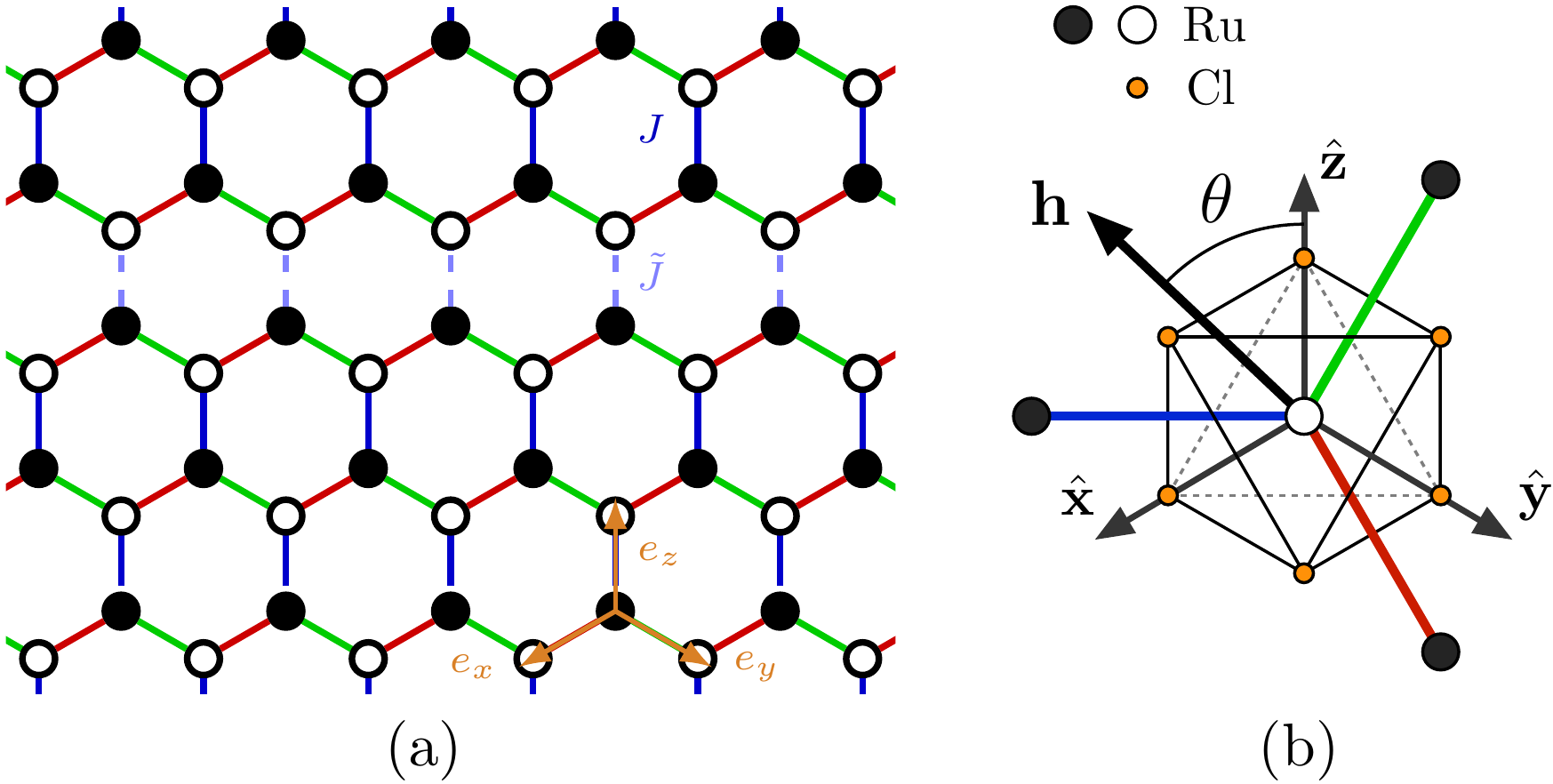}
    \caption{Kitaev honeycomb  model   with a line defect. Black and white circles represent the  sublattices.   Nearest-neighbor $x$-, $y$- and $z$-bonds   are colored in red, green and blue, respectively. (a) The   coupling  $\tilde J$   along the defect is represented by dashed lines. The vectors $ {\mathbf e}_{x,y,z}$    indicate the bond directions in the plane.   (b) In $\alpha$-RuCl$_3$, the Cartesian axes  with unit vectors $\hat {\mathbf x}$, $\hat {\mathbf y}$ and $\hat {\mathbf z}$ are defined by the  vertices of the ligand octahedra. The magnetic field forms an angle $\theta$ with $\hat {\mathbf z}$.    }
    \label{fig1}%
\end{figure} 

We model the 1D  defect as a line of weaker exchange bonds as shown in Fig. \ref{fig1}(a). For defect interaction $\tilde J=0$, the system reduces to two decoupled  Kitaev spin liquids with zigzag edges. 
 In the  non-Abelian  phase, the decoupled edges  harbor gapless  chiral Majorana modes. Aasen {\it et al.} \cite{Aasen2020} noted  that there  is a  critical  value  of  the    interaction    below  which  these 1D modes remain gapless.  The  reason  is  that   the  leading   interaction between  emergent  Majorana  fermions  across the interface   is irrelevant in the renormalization group sense. In addition to the effective field theory, the problem of seaming two Kitaev spin liquids was analyzed in Ref.  \cite{Aasen2020} by analogy with 1D lattice models that exhibit a transition 
 in the same universality class \cite{Rahmani_Affleck2015}. Here we start from the Kitaev    model and calculate the spectrum using a self-consistent mean-field approximation for the   interaction along the defect. Our approach reveals that the critical coupling  stems from a competition between this  interaction  and the Zeeman coupling for the dangling-bond spins.  

Below the critical coupling, the gapless  Majorana modes along the line defect can dominate the low-energy behavior of local response functions. To illustrate this point, we calculate the spin-lattice relaxation rate $1/T_1$  within the effective field theory. We find $1/T_1\propto T^3$ at low temperatures, in clear contrast with the exponential dependence expected for a gapped spin liquid. Remarkably, the cubic temperature dependence matches the result of the nuclear magnetic resonance (NMR) experiment in Ref. \cite{Zheng-gapless2017}. We then propose that the contribution from gapless 1D modes in samples with a low but finite  density of line defects might explain the discrepancy with other NMR experiments  that  observed a spin gap inside the putative Kitaev spin liquid  phase \cite{Baek2017,Jansa2018,Nagai2020}.

\emph{Microscopic model.---}Our starting point is the  spin-$1/2$ Kitaev honeycomb  model in a magnetic field \cite{Kitaev2006}: \be
H=-\sum_{\langle j,k\rangle_\gamma}J_{jk}\sigma_j^\gamma\sigma_k^\gamma -\mathbf h\cdot \sum_j\boldsymbol \sigma_j. \label{model}
\ee
Here  $\boldsymbol \sigma_j$ is the vector of Pauli operators at site $j$.  The Kitaev coupling    on nearest-neighbor bonds of type $\gamma=x,y,z$ takes the value  $J_{jk}=J$ in the bulk and $J_{jk}=\tilde J\ll J$ for the $z$ bonds along the line defect, see  Fig. \ref{fig1}(a). The nearest-neighbor vectors are $ {\mathbf e}_{x}=(-\frac{1}2,-\frac1{2\sqrt3})$, $ {\mathbf e}_{y}=(\frac{1}2,-\frac1{2\sqrt3})$ and  $ {\mathbf e}_{z}=(0,\frac1{\sqrt3})$, where we set the lattice parameter to unity.   The components of the magnetic field  $\mathbf h=h_x\hat {\mathbf x}+h_y\hat {\mathbf y}+h_z\hat {\mathbf z}$ are defined with respect to the axes fixed by the edge-sharing octahedra structure of $\alpha$-RuCl$_3$ \cite{Winter2017,Takagi2019}; see Fig. \ref{fig1}(b). Note that the $z$ axis is perpendicular to the $z$ bond. Importantly, even a small increase in the bond length  across the defect can significantly  suppress  the Kitaev coupling  \cite{Yadav2018}. Since the weaker coupling is the mechanism behind the persistence of gapless 1D modes, we consider an infinite line defect  without specifying its detailed properties, {\it e.g.}, the Burgers vector of dislocations   \cite{Petrova2013}.  For simplicity,  we neglect interactions beyond the pure Kitaev model \cite{Rau2014,Gordon2019}, which can renormalize the critical coupling  discussed in the following, but do not change    qualitative features of the transition or the temperature dependence of $T_1^{-1}$.

For $\mathbf h=0$, the Kitaev model is  solved by the representation $\sigma_j^\gamma=ib_j^\gamma c_{j}^{\phantom\gamma}$, where $b_j^\gamma$ and $c_j^{\phantom\gamma}$ are Majorana fermions \cite{Kitaev2006}. To restrict to the physical spin-$1/2$ Hilbert space, one imposes the local constraint $b_j^x b_j^yb_j^z c_{j}^{\phantom\gamma} =1$ for all sites. There is one conserved quantity for each hexagonal plaquette $p$, given by   $W_{p}  =  \prod_{\langle j,k \rangle \in \partial p} u_{jk}$, where  $ \quad u_{jk} = \dot{\imath} b_{j}^{\gamma} b_{k}^{\gamma} $   acts as a $\mathbb Z_2$ gauge field   on the $\langle j,k\rangle_\gamma$ bond  with site $j$ in sublattice A (black circles in Fig. \ref{fig1}) and site $k$ in sublattice B (white circles).

To make progress analytically, we follow Ref. \cite{Kitaev2006} and replace the Zeeman coupling in the bulk by a three-spin interaction generated by perturbation theory in the magnetic field. This approach  is justified by  a projection onto the low-energy sector where $W_p=1$ $\forall p$, which contains the exact ground state for $\mathbf h=0$.  In this sector, we can set $u_{jk}= 1$, freezing out all $b^\gamma$ fermions in the bulk. Moreover, the  three-spin interaction  effectively gaps out the spectrum of $c$ fermions with a topologically nontrivial mass, which is the main effect of time-reversal-symmetry breaking that we wish to capture with our model to describe the non-Abelian spin liquid phase. 
  On the other hand, for $\tilde J=0$, the $b^z$ fermions   associated with   the broken $z$ bonds, see Fig. \ref{fig1}(a), couple only to the $c$ fermions through  the  Zeeman  term proportional to $h_z$. In fact, for $\tilde J=0$ there is no energy cost for changing $W_p$ on plaquettes along the line defect.  On defect  sites we can still integrate out  the $b^{x,y}$ fermions, since these involve bonds with strong coupling $J$, but  the  $b^z$ fermions remain  dynamic at low energies.  For this reason,  we shall keep the projected Zeeman term   on defect sites.  As noted in  Ref. \cite{Kitaev2006}, without this term the   $b^z$    fermions at an edge would  decouple from the rest of the system and form a zero-energy flat band. The resulting  Hamiltonian in the regime $|\mathbf h|,\tilde J\ll J$ is\be
H_{\rm eff}=H_0-\tilde J\sum_{j\in \ell_1 } \sigma_j^z\sigma_{j+ {\mathbf e}_z}^z-h_z\sum_{j\in \ell_1\cup \ell_2}\sigma_j^z,
\ee
where $H_0=-J\sum_{ \langle j,k \rangle_{\gamma} }    \sigma_{j}^{\gamma} \sigma_{k}^{\gamma}    -  \kappa   \sum_{ \,  \langle j,k \rangle_{\alpha} , \langle k,l \rangle_{\beta} }  \sigma_{j}^{\alpha} \sigma_{k}^{\gamma} \sigma_{l}^{\beta}$ contains the standard bulk interactions  \cite{Kitaev2006} and $\ell_1$ and $\ell_2$ refer to the lines  of defect sites in   A and B sublattices, respectively. The coupling constant of the three-spin interaction is related to the magnetic field by $\kappa \propto h_xh_yh_z/J^2$.  Once we fix $u_{jk}=1$ in the bulk, $H_0$ becomes a quadratic  Hamiltonian for the $c$ fermions with nearest-  and next-nearest-neighbor couplings. In terms of Majorana fermions, we obtain
\be
H_{\rm eff}=H_0 
        - \tilde{J} \sum_{j \in \ell_1}   b^z_{j} b^z_{j+ {\mathbf e}_z} c^{\phantom z}_{j} c^{\phantom z}_{j+ {\mathbf e}_z}
        - \dot{\imath} h_{z}  \hspace*{-3mm} \sum_{j \in \ell_1  \cup \ell_2}   b_{j}^z c^{\phantom z}_{j} .\label{effHMajorana}
\ee

\emph{Mean-field theory.---}The model in Eq. (\ref{effHMajorana}) is not exactly solvable when both $\tilde J$ and $h_z$ are nonzero. While the  Zeeman term is quadratic, the hybridization of $b^z$ and $c$  spoils the conservation of $\dot{\imath} b^z_{j} b^z_{k}$ on   defect bonds. Here we use a Majorana mean-field approximation for the quartic term. Similar  approaches  have been shown to capture  phase transitions driven by integrability-breaking bulk interactions in the extended Kitaev model \cite{Nasu2018,Knolle2018}. We adopt the mean-field parameters $ \chi_b =  \left\langle  \dot{\imath}  b^{z}_{j}b_{j+\mathbf e_z}^{z}  \right\rangle$ and  $\chi_c  =  \left\langle   \dot{\imath} c_{j} c_{j+\mathbf e_z}  \right\rangle$ for $j\in\ell_1$. Performing a mean-field decoupling of the interaction $V= \tilde{J} \sum_{j \in \ell_1}  (\dot{\imath} b^z_{j} b^z_{j+\mathbf e_z})(\dot{\imath} c^{\phantom z}_{j} c^{\phantom z}_{j+\mathbf e_z})$, we obtain
\begin{equation}
     V_{\text{MF}} =  \tilde{J} \, \sum_{j \in \ell_1} \left( \, \dot{\imath} \chi_c  b^{z}_{j}b_{j+\mathbf e_z}^{z} + \dot{\imath} \chi_b c_{j}c_{j+\mathbf e_z}   - \chi_b\chi_c \, \right)\; . 
\label{eq:4-MF-H}
\end{equation}
The replacement of  $V$ by $V_{\rm MF}$   in Eq. (\ref{effHMajorana}) yields the mean-field Hamiltonian  $H_{\rm MF}$.

We diagonalize the mean-field Hamiltonian numerically on a finite system with periodic boundary conditions. The geometry can be viewed as a torus with length $L_x$  in the direction parallel to the line defect, along  which the  model has translational invariance,  and containing $L_y$ sites in the transverse direction.  Representing a site by  a pair of coordinates $j=(x,y)$, we define the Fourier transformed   fermions \be
d_{q,y}=\frac1{\sqrt{2L_x}}\sum_{x=1}^{L_x} e^{-iqx}d_{x,y},\quad y=0,\dots,L_y+1,
\ee  
where $d^{\phantom z}_{x,y}=c^{\phantom z}_{x,y}$ for $y=1,\dots,L_y$, $d^{\phantom z}_{x,0}=b^{  z}_{x,1}$ and $d^{\phantom z}_{x,L_y+1}=b^{  z}_{x,L_y}$.  In this notation, $y=1$ and $y=L_y$ correspond to    lines  $\ell_1$ and $\ell_2$, respectively.   The mean-field Hamiltonian is quadratic in the complex fermions $d_{q,y}$ and can be cast in the form\be
 H_{\text{MF}} =
     \sum_{0<q \leq \pi } \sum_{y,y'} \dot{\imath} A_{y,y'}(q) d_{q,y}^{\dagger} d_{q,y'}^{\phantom{\dagger}}.\label{HMF}
\ee
Thus, the problem reduces to diagonalizing  the Hermitean matrix $\dot{\imath} A(q) $ of dimension $L_y+2$ whose components are given in the Supplemental Material \cite{SM}. The normal modes are given by $\gamma^\dagger_{q,n}=\sum_{y}U_{y,n}(q)d^\dagger_{q,y}$, where the unitary matrix $U(q)$  depends on  $\chi_b$ and $\chi_c$. 

The mean-field parameters  must be determined by self consistency of the approximation. We obtain the self-consistency equations by calculating $\chi_b$ and $\chi_c$ as expectation values in the mean-field ground state, expressed in terms of    the matrix elements   $U_{y,n}(q)$. To account for the magnetic-field dependence of $\kappa$, we set $\kappa=h_xh_yh_z/J^2$.  In addition, we parametrize the field direction by polar and azimuthal angles $\theta$ and $\phi$ with respect to the  axes in Fig. \ref{fig1}(b).  The mean-field parameters  are then real functions of $\tilde J/J$, $|\mathbf h|/J$, $\theta$ and $\phi$.


\begin{figure}[t]
\includegraphics[width=.95\columnwidth]{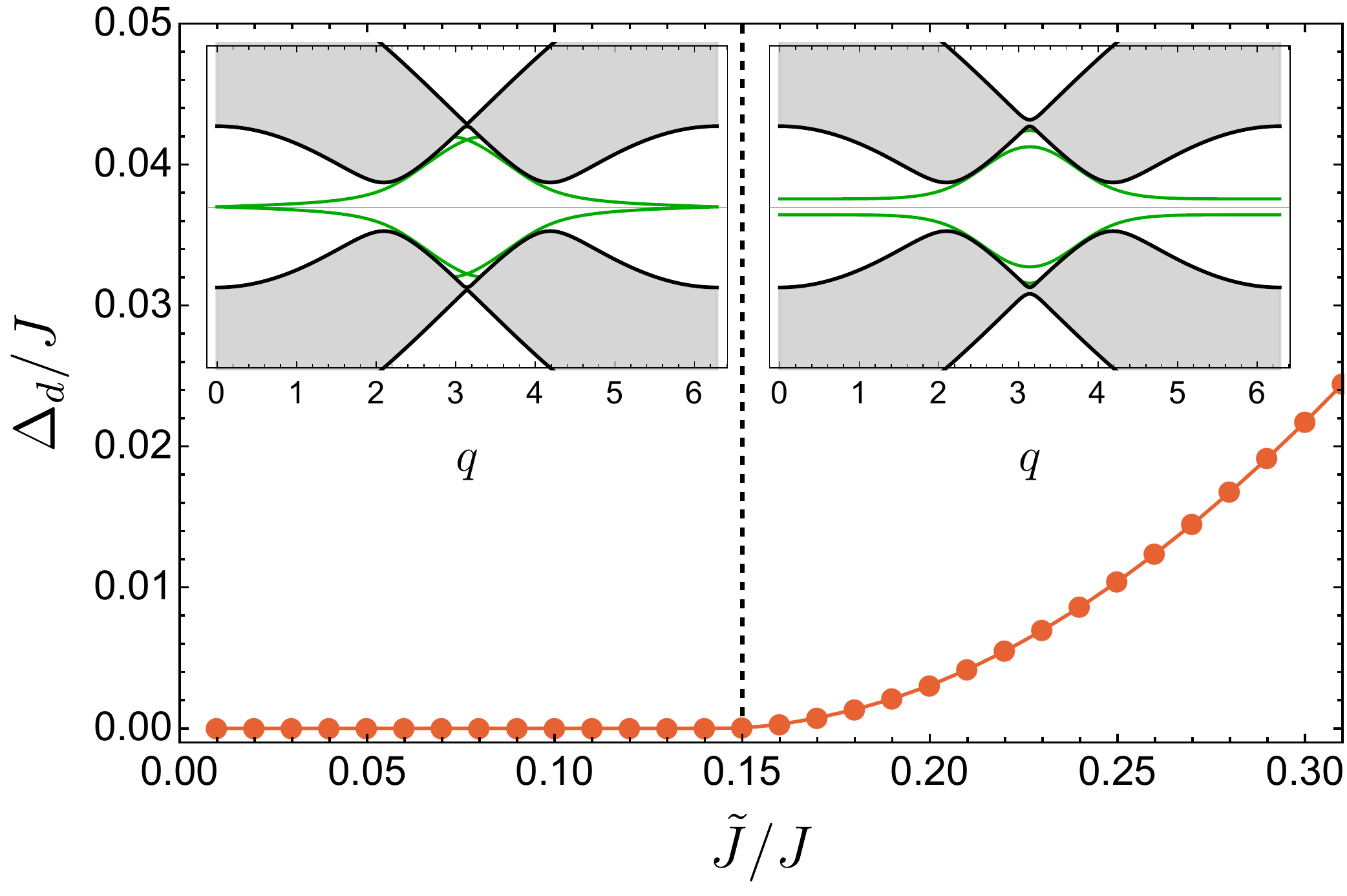}
\caption{Energy gap   for  the Majorana modes  bound to the line defect as a function of the modified coupling $\tilde J$. Here we fix the magnetic field along the [111] direction with $|\mathbf h|=0.8J$.  The dashed line indicates the critical coupling $\tilde{J}_c$. The inset shows the spectrum below and above $\tilde J_c$. The green  lines refer to  defect modes  and the  continuum   to gapped bulk  modes. 
}
\label{fig2}
\end{figure}

Our numerical  results confirm  that the mean-field parameters   vanish below a  critical coupling $\tilde J_c>0$. In this case, the two sides of the line defect remain decoupled and the spectrum exhibits chiral Majorana modes with linear dispersion near $q=0$. For $\tilde J>\tilde J_c $, we find that both  $\chi_b$ and $\chi_c$ become nonzero with a continuous transition, in contrast with the first-order transition obtained by a variational analysis of the continuum model in Ref. \cite{Aasen2020}. As a check of our approach, we observe that $\chi_b\to 1$   in the limit $|h_z|/ \tilde J\to 0$, as expected since $ib_j^zb_k^z$ in Eq. (\ref{effHMajorana}) become conserved $\mathbb Z_2$ operators when we neglect the Zeeman coupling.   For nonzero $\chi_b$ and $\chi_c$, the Majorana fermions on different sides hybridize and the 1D mode acquires a gap $\Delta_d$.    Figure \ref{fig2} shows the gap for a magnetic field along the [111] direction. Interestingly,   for a wide range of $\tilde J >\tilde J_c$ the  gap  $\Delta_d$ remains much smaller than the bulk gap $\Delta_b=6\sqrt3 \kappa$.

\begin{figure}[t]
\centering
\includegraphics[width=.95\columnwidth]{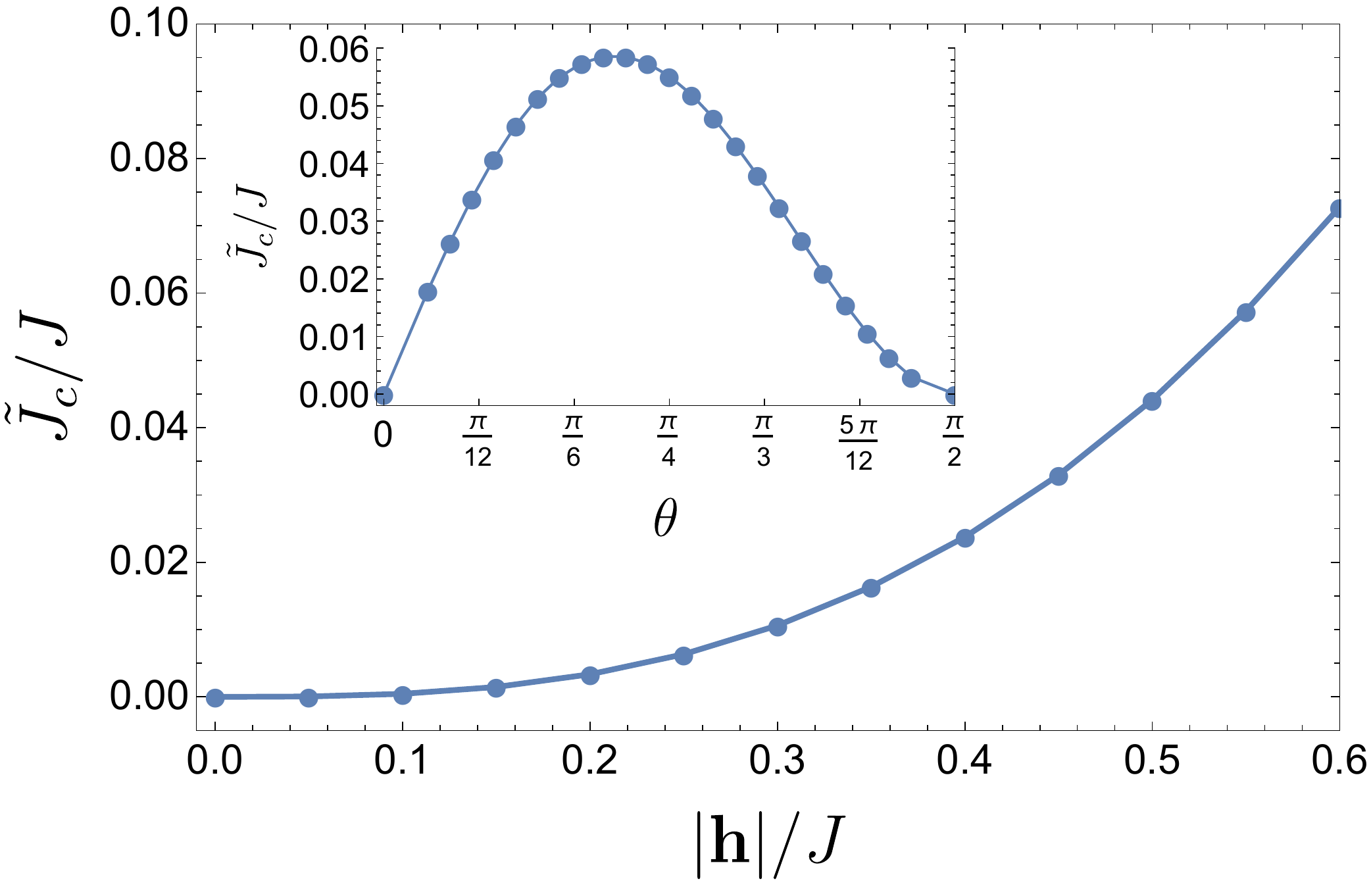}
\caption{Critical value  of the interaction at the line defect as a function of the magnetic field.  The field is fixed along the $[111]$ direction ($\theta = \tan^{-1}\sqrt{2}$ and $\phi = \pi/4$). Inset:  critical coupling as a function of $\theta$ for fixed $|\mathbf h|=0.5J$ and $\phi=\pi/4$.   }
\label{fig:jc-h.}
\end{figure}

While the gap goes smoothly to zero at the transition, we can determine the critical point  precisely by expanding the self-consistency equations for small values of the mean-field parameters. To first order in $\chi_b$ and $\chi_c$, the equations  take the form $\chi=\tilde J \Gamma\chi$, where $\chi=(\chi_b,\chi_c)^t$ and $\Gamma$ is a $2\times2$ matrix easily computed in terms of the unitary matrix $U^{(0)}(q)$ that diagonalizes the mean-field Hamiltonian for $\chi_b=\chi_c=0$ \cite{SM}. Requiring a nontrivial solution to the linear equation with $\tilde J\to \tilde J_c$,  we obtain   $\tilde J_c=2(\text{tr}\,\Gamma)^{-1}[-w+\sqrt{w(w+1)}]$ with $w=(\text{tr}\,\Gamma)^2/(4|\det\Gamma|)$. The dependence of $\tilde{J}_c$ on the magnetic field is shown in Fig. \ref{fig:jc-h.}.  For fixed field direction, we  observe a power-law behavior $\tilde{J}_c \sim \vert \mathbf{h} \vert^{\beta}$ with exponent $\beta \approx  3$. Moreover, the critical coupling  varies with the field direction through the dependence on 
$\kappa$ in the bulk and $h_z$  at the line defect. Like the bulk gap,  $\tilde J_c$ vanishes when   any of  the components $ h_x$, $h_y$ or $h_z$ go to zero.  We have verified that the critical coupling  also remains finite for in-plane fields, except for the special directions in which the bulk gap closes \cite{SM}.

\emph{Low-energy effective theory.---}For  weak coupling   ${\tilde J<\tilde J_c}$, the low-energy sector is described by two chiral Majorana fermions, each   associated with one side of the line defect. To derive the effective field theory, we  expand the matrix $A(q)$ in Eq. (\ref{HMF}) to first order in $q$. The bound  state wave functions decay exponentially with the distance  from the  defect and can be determined exactly  for $q\to0$. Solving the eigenvalue equation with dispersion $\varepsilon_{R/L}(q)=\pm vq$ for the chiral modes, we find an analytical expression for  the velocity    \cite{SM}\be
v=\frac{|\kappa| h^2_z}{J^2 +h^2_z/3}. 
\ee
Note that the velocity vanishes  when $h_z\to0$ or $\kappa\to0$. 

The low-energy  Hamiltonian is effectively 1D:
\be
    H_{\rm low}  = \sum_{q>0} vq \left( \gamma^{\dagger}_{qR}\gamma^{\phantom{\dagger}}_{qR} - \gamma^{\dagger}_{qL}\gamma^{\phantom{\dagger}}_{qL} \right)  ,
\ee
where $\gamma_{q\alpha}$, with $\alpha=R,L$, are the annihilation operators for  right and left movers. In the continuum limit, the   chiral Majorana fermions are $\gamma_\alpha(x)=\sqrt{2/L_x}\sum_{q>0}(e^{iqx}\gamma^{\phantom\dagger}_{q\alpha}+e^{-iqx}\gamma_{q\alpha}^\dagger)$. We can then write
\begin{equation}
    H_{\text{low}} = \frac{1}{4} \int dx \; \left[ \, \gamma_{R} (- \dot{\imath} v \partial_x ) \gamma_{R}  +  \gamma_{L} ( \dot{\imath} v \partial_x ) \gamma_{L} \, \right].\label{Hlow}
\end{equation}

Next, we  calculate the representation of the spin operator in the low-energy theory in terms of the chiral Majorana fermions. For the geometry in Fig. \ref{fig1}(a), only  the $\sigma_j^z$ component for sites $j$ near the line defect has a nonzero projection onto the gapless modes. Expanding $c_j$ and  $b_j^z$ on a defect site in terms of the normal modes to first order in $q$, we obtain\be
S_j^z=\frac12\sigma_j^z\sim S_\alpha^z(x)=\frac{i\alpha s}2 \gamma_\alpha (x)\partial_x\gamma_\alpha(x),\label{spinop}
\ee 
where $s=J^2 |\kappa| h_z/(J^2+h_z^2/3)^2$ and we select  $\alpha=R,L=+,-$ according to the chiral mode whose wave function lies on the same side as  site $j$. The   representation  in Eq. (\ref{spinop}) could be argued     on symmetry grounds  \cite{Aasen2020}.   We stress that, in contrast  with the usual  parton representation using  complex Abrikosov  fermions $\mathbf S_j=\frac12f_{a}^\dagger \boldsymbol \sigma_{ab} f_b$ \cite{Savary2016,wen2004quantum}, the  Majorana fermion representation requires  a spatial derivative, which increases the scaling dimension of the operator in Eq. (\ref{spinop}).      

The projection of the quartic term in Eq. (\ref{effHMajorana}) onto the gapless modes yields   $V\sim g\int dx\, \gamma_R\partial_x\gamma_R\gamma_L\partial_x\gamma_L$ with $g\sim \tilde J s^2$. This irrelevant interaction  is the leading perturbation to the low-energy  fixed-point Hamiltonian \cite{Rahmani_Affleck2015,Aasen2020}. Crucially, the mass term $im\gamma_R\gamma_L$ is forbidden, as   local operators  must be   bilinears of the emergent Majorana fermions on the same side of the line defect. As long as $v>0$, the transition occurs at a finite critical coupling,    spontaneously breaking the $\mathbb{Z}_2\times \mathbb{Z}_2$ symmetry of independently flipping the signs of $\gamma_R$ and $\gamma_L$. Beyond the mean-field level, the critical point is described by the tricritical Ising conformal field theory \cite{Rahmani_Affleck2015}.

\emph{NMR response.---}The  effective field theory allows us to calculate the spin-lattice relaxation rate  at low temperatures  for a nuclear spin adjacent to the line defect. When restricted to the contribution from the gapless 1D mode,  the linear  response formula for $T_1^{-1}$  becomes  \cite{Carretta2011,Sirker2011} 
\begin{equation}
    \frac{1}{T_1}  =  \frac{\gamma_{N}^2}{ 2 L_x }    \sum_{q} \, \vert A_{\text{hf}}(q) \vert^2  S(q,\omega_0)  , \label{eq:6-t1}
\end{equation}
where $\gamma_{\text{N}}$ is the nuclear gyromagnetic ratio for   $^{35}$Cl NMR in $\alpha$-RuCl$_3$ \cite{Zheng-gapless2017,Baek2017,Jansa2018,Nagai2020}, $A_{\text{hf}}(q)$ is the hyperfine coupling form factor, $\omega_{0} = \gamma_{\text{N}}\,  \vert \bf{h} \vert $ is the Larmor nuclear resonance frequency, and 
\be
     S(q,\omega_0)  =   \int dx dt \, e^{\dot{\imath} (\omega_0 t-q x)}   \langle  \tilde{S}_{\alpha}^{+}(x,t) \tilde{S}_{\alpha}^{-}(0,0)      \rangle  \label{NMR}
\ee
is  the transverse dynamical spin structure factor at temperature $T$ for either value of $\alpha$.  Here $\tilde S^{\pm}(x,t)=\tilde S^x(x,t)\pm i \tilde S^y(x,t)$ are  time-evolved ladder operators   that  perform spin flips with respect to the magnetic-field axis. On the other hand, the spin components  $S^\gamma$  were originally defined with respect to the axes in Fig. \ref{fig1}(b).  Rotating the coordinate system, we obtain
\bea
         \tilde{S}^{\pm}    &=& - S^{z}  \sin \theta +S^{x}\left( \cos \theta \, \cos \phi \mp \dot{\imath} \sin \phi   \right)\nonumber  \\
         &&+ S^{y} \left( \cos \theta \, \cos \phi \pm \dot{\imath} \sin \phi   \right).
\eea
Since only the $S^z$ operator has a nonzero projection onto  the gapless modes, we have $ \langle  \tilde{S}_{\alpha}^{+}(x,t) \tilde{S}_{\alpha}^{-}(0,0)\rangle= \langle  {S}_{\alpha}^{z}(x,t) {S}_{\alpha}^{z}(0,0)\rangle \sin^2\theta $, with ${S}_{\alpha}^{z}(x)$ given by Eq. (\ref{spinop}).

We calculate $T_1^{-1}$ using Green's functions for noninteracting Majorana fermions  described  by the   Hamiltonian in Eq. (\ref{Hlow}).  The experimentally relevant regime with $T\sim 1$ K and $\omega_0\sim 10$ MHz  is $\omega_0\ll T$, where we set  $\hbar=k_B=1$.  In this regime, the   dynamical structure factor can be written as $S(q,\omega_0)\approx -(2T/\omega_0)\text{Im}\chi^{\rm ret}(q,\omega_0)$, where $\chi^{\rm ret}(q,\omega_0)$ is the retarded dynamical susceptibility for the $S_\alpha^z(x)$ operator. The latter can be calculated by analytical continuation of the Matsubara correlation function  \cite{SM}. We find    
\begin{equation}
    \dfrac{1}{T_1} \approx    \frac{  \pi^2 s^2\gamma_{\text{N}}^2}{6v^4}   \vert A_{\text{hf}}(0)\vert^2\sin^2 \theta \; T^3  .\label{1T1}
\end{equation}
This result is valid for $T\lesssim v$, since $v$ sets  the high-energy cutoff of the effective field theory  when the lattice parameter is  set to unity. Note that the temperature window shrinks to zero  for  $\kappa\to 0$.

A cubic temperature dependence in the spin-lattice relaxation rate has been observed experimentally  \cite{Zheng-gapless2017} and interpreted as evidence for a gapless spin liquid in $\alpha$-RuCl$_3$.   Indeed, $T_1^{-1}\sim T^3$ is expected for a generic   Kitaev spin liquid with a  massless Dirac spectrum  \cite{Song2016}, possible when the magnetic field points along the particular directions in which $\kappa=0$. However, the results of Ref.  \cite{Zheng-gapless2017}  showed a cubic temperature dependence over a broad field range, independent of  orientation.  For  magnetic fields above 12 T, the spin-lattice relaxation rate decays faster with  decreasing temperature. The deviation from the $T^3$ behavior  at high fields indicates a suppression of the mechanism responsible for the gapless modes, as expected when the material enters the trivial polarized phase.  Meanwhile, other  measurements of  $1/T_1$ in $\alpha$-RuCl$_3$   favor a picture of  fully gapped  spin excitations \cite{Jansa2018,Nagai2020}.

Here we suggest that the apparently gapless behavior  may have  its origin in  1D modes bound to line defects. Importantly,   the effective   field theory shows that the robustness of these   gapless   modes  is a \emph{universal} property  of the non-Abelian Kitaev spin liquid.  While   the   realistic spin model for $\alpha$-RuCl$_3$  must include   Heisenberg and off-diagonal  exchange interactions   \cite{Rau2014} neglected in Eq. (\ref{model}),   
our main conclusions do not depend on microscopic details. As long as the perturbations to the Kitaev model do not destroy the topological order,  they can only renormalize the prefactor of $1/T_1$  in Eq. (\ref{1T1}).  The $T^3$ dependence  only relies on the existence of  chiral Majorana modes with linear dispersion. In  fact, the exponent  can be traced back to the scaling dimension of  the spin operator in Eq. (\ref{spinop}). By contrast, for  a 1D system of complex fermions  described by Luttinger liquid theory, the small-$q$ contribution to the spin-lattice relaxation rate scales as  $T_1^{-1}\sim T$  \cite{Sirker2011,Sachdev1994}. Thus,  our result does not   follow  from a usual density-of-states factor,  but is connected with the Majorana fermion nature of the elementary excitations.  We  note that other types of defects in the Kitaev spin liquid, such as site  dilution, can also give rise to a  power-law dependence in $T_1^{-1}$, but with lower exponents  \cite{Nasu2021}.

Our results  also reveal a characteristic dependence on the magnetic-field direction.   However, the geometric factor $\sin^2 \theta$ holds only for a line defect running between zigzag edges. For  more general  geometries,   different spin components may have  projections  onto the chiral Majorana modes, modifying  this geometric factor.  For randomly oriented  line defects, the angular dependence  averages out, which is consistent with the  experiment  of Ref.  \cite{Zheng-gapless2017}.  An alternative explanation, put forward in Ref. \cite{Liu2018}, is that the intermediate phase of $\alpha$-RuCl$_3$ might be described by a  U(1) spin liquid whose gap remains small, below the measurement temperature,  for arbitrary field directions.  While the nature of this phase is  under scrutiny again \cite{Czajka2021}, our proposal highlights the role of line defects when unraveling the properties of Kitaev materials. To distinguish between different scenarios for the  gapless behavior, it would be interesting to single out the contribution from line defects by controlling their density and orientation \cite{Komsa2013} in different samples.

\emph{Conclusions.---}  We showed that gapless Majorana modes bound to line defects can survive in the bulk of non-Abelian Kitaev spin  liquids, with clear signatures in  low-energy properties. As an example, we showed that these  modes give rise to a cubic temperature dependence of the spin-lattice relaxation rate, offering an  explanation for the experimental findings of Ref. \cite{Zheng-gapless2017}. The critical value of the interaction below which the 1D  modes remain gapless   can be tuned by the magnitude and orientation of the external magnetic field.

\begin{acknowledgments}
We thank R. Egger and E. Miranda for helpful discussions. We  acknowledge funding by      Brazilian agencies CAPES (L.R.D.F.) and  CNPq (R.G.P.). Research at IIP-UFRN is supported by Brazilian ministries MEC and MCTI. This work was also  supported by  a grant from Associa\c{c}\~ao Instituto Internacional de F\'isica. 
\end{acknowledgments}

\bibliographystyle{apsrev4-1}       
\bibliography{ref}

\onecolumngrid

\appendix
 
\section{Supplemental Material:  Gapless excitations in non-Abelian Kitaev spin liquids with line defects}
  

\subsection{1. Self-consistent mean-field approach}
In the mean-field Hamiltonian written in  Eq. (6) of the main text,     the Hermitean   matrix of dimension $L_y+2$ is
\begin{equation}
  \im A(q) = 
   \left(
\begin{array}{cccccccccccc}
 0 & \im\gamma & 0 & 0 & 0 & 0 & 0 & 0 & 0 & \im\tilde{J}\chi_c \\
-\im \gamma  & \alpha  & \im s & -\beta  & 0 & 0 & 0 & 0 & \im \tilde{J} \chi_b & 0  \\
 0 & -\im  s & -\alpha  & \im r & \beta  & 0 & 0 & 0 & 0 &0 \\
 0 & -\beta  & \im r & \alpha  & i s & -\beta  & 0 & 0 & 0 & 0 \\
 0 & 0 & \beta  & -\im s & -\alpha  & \im r & \beta  & 0 & 0 & 0  \\
 0 & 0 & 0 & \ddots  &  \ddots & \ddots  & \ddots & \ddots  & 0 & 0 \\
 0 & 0 & 0 & 0 & \ddots & \ddots & \ddots  & \ddots &  \ddots  & 0  \\
0 & 0 & 0 & 0 & 0 & -\beta  & -\im r & \alpha  & \im s & 0 \\
0 &  -\im \tilde{J} \chi_b & 0 & 0 & 0 & 0 & \beta  & -\im s & -\alpha  & -\im \gamma  \\
 -\im \tilde{J} \chi_c & 0 & 0 & 0 & 0 & 0 & 0 & 0 & \im \gamma  & 0 \\
\end{array} \right), \label{eq:SM-iA}
\end{equation}
with matrix elements as in Ref. \cite{Kitaev2006}:
\begin{align}
    \alpha       & = 2  \kappa    \sin(q)   ,  \\
    \beta        & = 2    \kappa    \sin(q/2)   ,  \\
    r               & =   J     ,   \\
    s               & = -2     J   \cos(q/2)     , \\  
    \gamma          &   = - h_{z}  .
\end{align}
The mean-field parameters can be written as
\begin{align}
\chi_b &=  \left\langle  \dot{\imath}  d_{x,0}d_{x,L_y+1}  \right\rangle   , \\
\chi_c &=  \left\langle  \dot{\imath}  d_{x,1}d_{x,L_y}  \right\rangle ,    
\end{align} 
where the expectation values are calculated in the mean-field ground state. 

\begin{figure}[t]
    \centering
    \includegraphics[width = .45\textwidth]{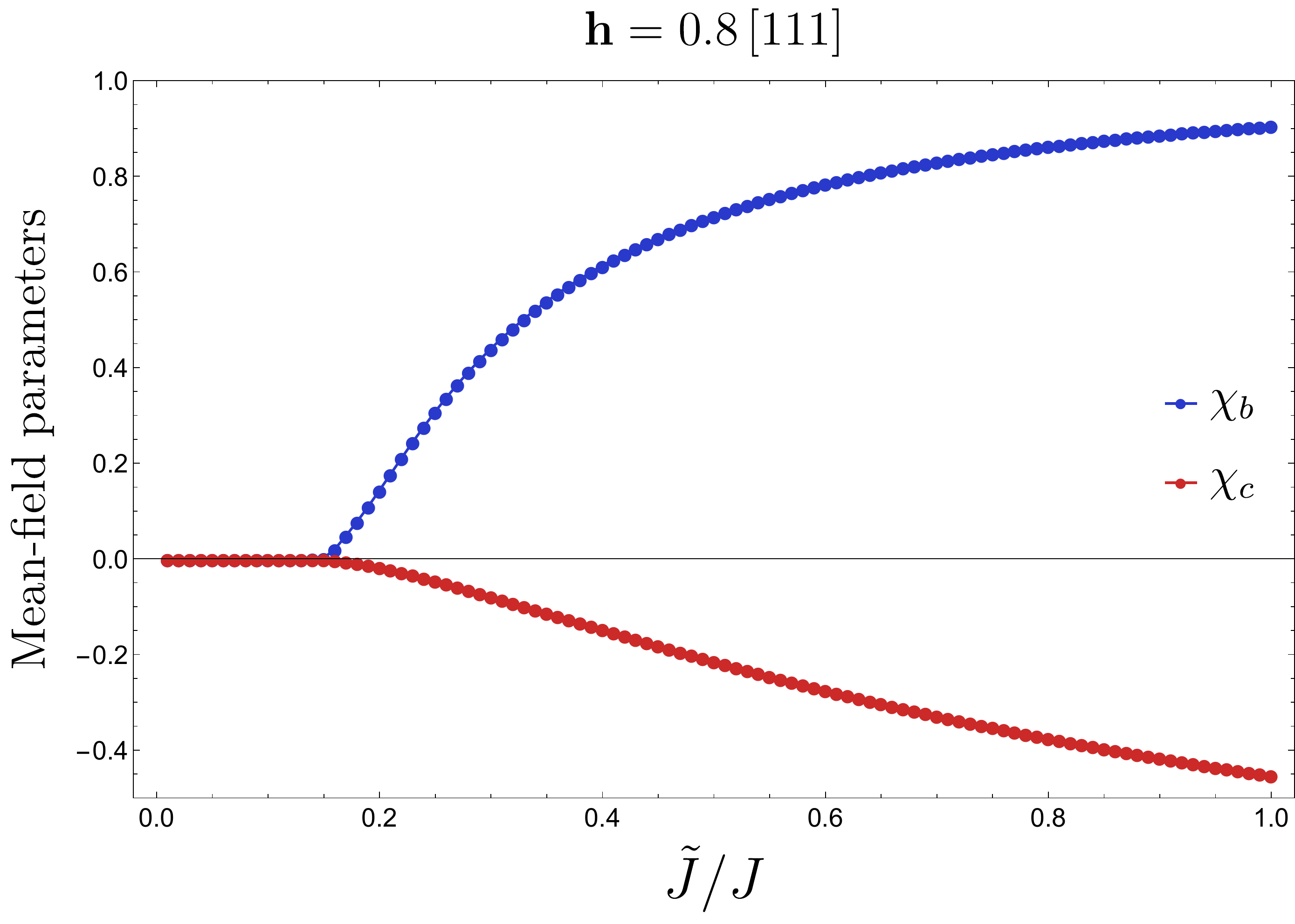}\qquad \qquad 
    \includegraphics[width = .45\textwidth]{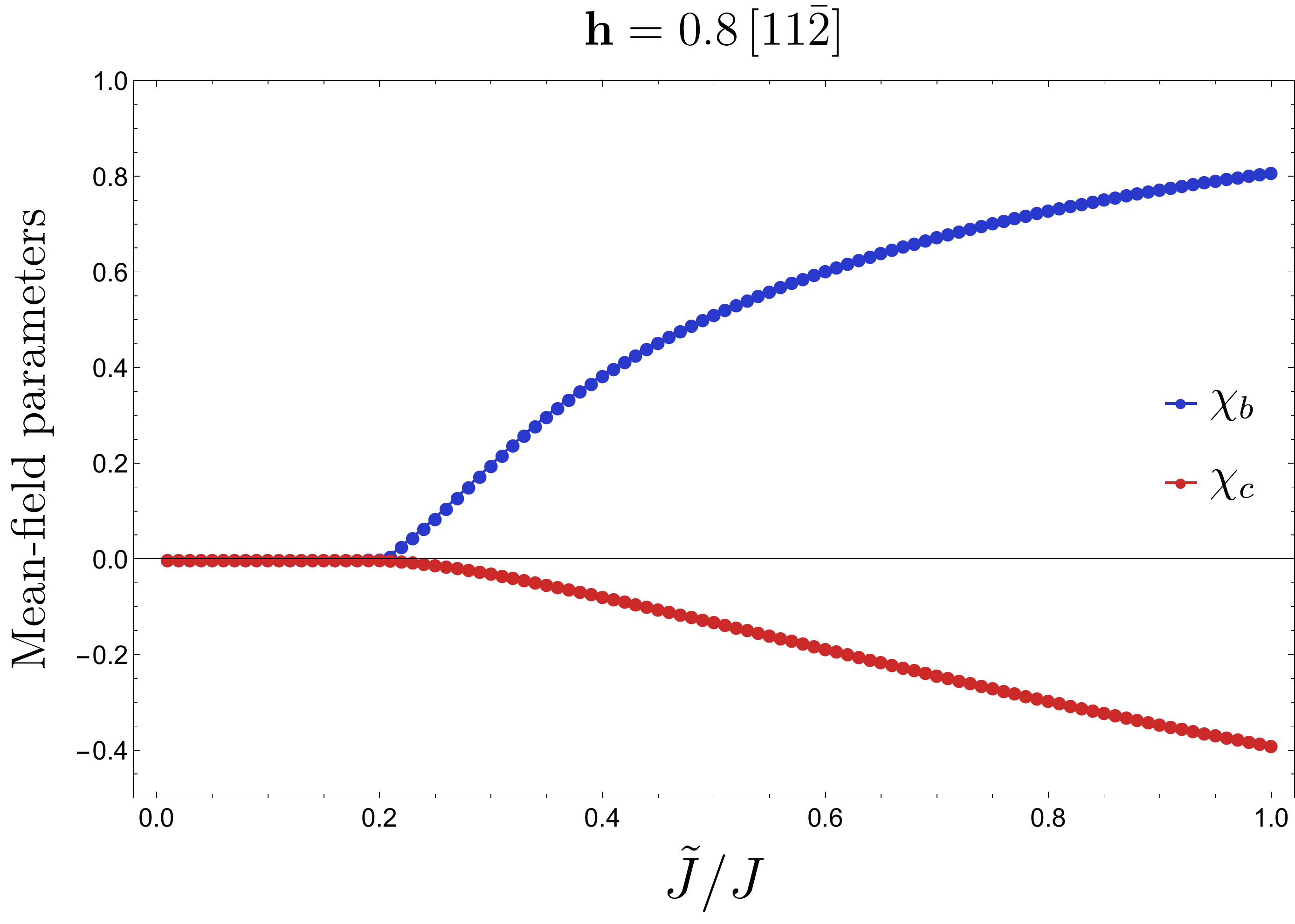}
    \caption{Mean-field parameters  for a magnetic field  with  $\vert \mathbf{h} \vert = 0.8 J$ and two representative directions. Left:  magnetic field in the [111] direction, perpendicular to the $ab$ plane which contains the honeycomb lattice. Right:  in-plane field in the [$11\bar 2$] direction, usually denoted as the $a$ axis. }
    \label{fig:SM-f4}
\end{figure}

Performing  a unitary transformation \begin{align}
    d_{q,y}^{\phantom{\dagger}}  &=  \sum_{n=1}^{L_y+2}   U_{y,n}(q)   \gamma_{q,n}   ,\label{eq:SM-d-gamma}
\end{align} 
we diagonalize the mean-field Hamiltonian in the form
\be  H_{\text{MF}} = \sum_{0<q\leq\pi }  \sum_{n=1}^{L_y+2} \varepsilon_{n}(q) \gamma_{q,n}^{\dagger} \gamma_{q,n}   .
\ee 
The eigenvalues are sorted such that 
$ \varepsilon_{n + \frac{L_y}{2}}(q)  =  -  \varepsilon_{n}(q) \leq 0 $, for  $ 1\leq n \leq  \frac{L_y}{2}+1$.  The mean-field ground state is a Dirac sea in which  all negative-energy states are occupied, 
so that  $\langle \gamma^{\dagger}_{q,n} \gamma_{q',n'}^{\phantom{\dagger}} \rangle   =   \delta_{qq'}\delta_{nn'}\Theta(-\varepsilon_{n}(q))$. 
As a result,  we can express the mean-field parameters in terms of the matrix elements $U_{y,n}(q)$ as follows:
\begin{align}
        \chi_b   &= \frac{4}{ L_x} \sum_{0 < q \leq \pi}  \sum_{n=1}^{\frac{ L_y}{2}+1}   \text{Im} \left[   U_{0,n}^{\ast}(q)U_{ L_y+1,n}(q)   \right]   , \label{eq:SM-chi-b}   \\
        \chi_c   &=  \frac{4}{ L_x}\sum_{0 < q \leq \pi}  \sum_{n=1}^{\frac{ L_y}{2}+1}   \text{Im} \left[   U_{1,n}^{\ast}(q)U_{ L_y,n} (q)  \right]    . \label{eq:SM-chi-c}
\end{align}
In Fig. \ref{fig:SM-f4}, we show the result for the mean-field parameters obtained by numerically iterating Eqs. (\ref{eq:SM-chi-b}) and (\ref{eq:SM-chi-c}). Note that $\chi_b$ and $\chi_c$ vanish below the critical coupling $\tilde J_c$. Moreover, for a fixed value  of $|\mathbf h|$, the critical coupling  for an in-plane field along the $a$ axis  is slightly   larger than for a field perpendicular to the $ab$ plane. 

\subsection{2. Determining the critical coupling}

Here we will use perturbation theory for small values of the mean-field parameters to calculate precisely the critical coupling $\tilde{J}_c$.  
Let $\{ \phi_{n}(q) \}$  be  the eigenvectors  of $A(q)$, represented as  column vectors and corresponding to the columns of the matrix $U_{y,n}(q)$. These vectors satisfy the eigenvalue equation 
\begin{equation}
    \im A(q)    \phi_{n}(q)       =       \varepsilon_{n}(q)    \phi_{n}(q)   .
\end{equation} 
Close to the critical point, we may expand this equation linearly in $\chi_b$ and $\chi_c$, which    appear in $A(q)$ multiplying  $\tilde{J}$. We define the coefficients of the expansion as
\begin{align}
    \phi_{n}(q) &=  \phi_{n}^{(0)}(q)  +  \tilde{J} \, \chi_b  \,  \phi_{n}^{(1b)}(q)   +   \tilde{J} \, \chi_c \,   \phi_{n}^{(1c)}(q) +...   , \label{eq:SM-phi-exp} \\[5pt]
    \varepsilon_{n}(q)   &=    \varepsilon_{n}^{(0)}(q)   +   \tilde{J} \, \chi_b   \, \varepsilon_{n}^{(1b)}(q)   +   \tilde{J} \, \chi_c \,   \varepsilon_{n}^{(1c)}(q) +...   , \\[5pt]
     \im A(q)  &=   \im A^{(0)}(q)   +    \tilde{J} \, \chi_b \, \mathcal{V}_b    +    \tilde{J} \, \chi_c \, \mathcal{V}_c+ ...,
\end{align}
where the matrices are
\begin{equation}
  \mathcal{V}_b   =   \dot{\imath}     
   \left(
\begin{array}{cccccccc}
 0 & 0 & 0 & 0 & 0 & 0  \\
0&0&0&0& 1 & 0 \\
 0 &  \ddots  &  \ddots & \ddots    & 0 & 0  \\
 0 & 0 &  \ddots & \ddots & \ddots    & 0  \\
0 &  -1 &0&0&0&0  \\ 
0 & 0&0&0&0&0 \\
\end{array} \right) ,  \qquad \qquad   
  \mathcal{V}_c   =   \dot{\imath}    
   \left(
\begin{array}{cccccccc}
 0 & 0 & 0 & 0 & 0  &  1 \\
0&0&0&0& 0 & 0 \\
 0 &  \ddots  &  \ddots & \ddots   & 0 & 0  \\
 0 & 0 &  \ddots & \ddots & \ddots   & 0  \\
0 &  0 &0&0&0&0  \\ 
-1 & 0&0&0&0&0 \\
\end{array} \right)   .
\end{equation}
The coefficients in the first-order correction to the energies can be calculated   as
\be      \varepsilon_{n}^{(1b)}(q)    =    [ \phi_{n}^{(0)}(q)  ]^t  \mathcal{V}_b  \phi_{n}^{(0)} (q)    , \qquad \qquad
      \varepsilon_{n}^{(1c)}(q)    =   [ \phi_{n}^{(0)} (q)]^t     \mathcal{V}_c  \phi_{n}^{(0)}(q)   . \ee


The unperturbed  eigenvectors $\phi_{n}^{(0)}(q)$  are given by the column vectors of the matrix  $U^{(0)}(q)$, which we obtain numerically by diagonalizing the Hamiltonian with $\tilde{J}=0$. 
To  first order in $\chi_b$ and $\chi_c$, we obtain  
\begin{equation}
        U_{y,n}(q)   =   U_{y,n}^{(0)}(q) + \tilde{J} \sum_{n' \neq n} \left\{   \chi_b  \frac{ [ \phi_{n}^{(0)} (q)]^t     \mathcal{V}_b  \phi_{n'}^{(0)}(q)   }{\varepsilon^{(0)}_{n}(q) - \varepsilon^{(0)}_{n'}(q)   } + \chi_c \frac{  [ \phi_{n}^{(0)}(q)]^t       \mathcal{V}_c  \phi_{n'}^{(0)}(q)   }{\varepsilon^{(0)}_{n}(q) - \varepsilon^{(0)}_{n'} (q)  }    \right\} U_{y,n'}^{(0)}(q)   .
\end{equation}

 Equations (\ref{eq:SM-chi-b}) and (\ref{eq:SM-chi-c}) involve   the imaginary part of the product  $U_{y,n}^{\ast}(q)U_{y',n}(q)$. The latter vanishes to zeroth order in mean-field parameters. Expanding to first order, we obtain the linear equations 
\be \left( \begin{array}{c}
             \chi_b  \\
             \chi_c 
        \end{array}\right) =   \tilde{J} \,
        \left( \begin{array}{cc}
             \, \Gamma^{bb}  &    \, \Gamma^{bc} \\
                \, \Gamma^{cb}  &   \,\Gamma^{cc} 
        \end{array} \right)
        \left( \begin{array}{c}
             \chi_b  \\
             \chi_c 
        \end{array}\right) + ..., \label{eq:SM-chi-Gamma}
\ee
with coefficients
\begin{align}
    \Gamma^{bb} 
      &=     \Gamma^{cc} 
      =    \frac{4}{ L_x}\text{Re}\left[ \sum_{0<q\leq \pi}
    \sum_{n,n'}' \frac{1 }{\varepsilon^{(0)}_{n} - \varepsilon^{(0)}_{n'}}  \left( U_{ 1,n }^{(0)\ast}    U_{   L_y,n' }^{(0)}   -  U_{  1,n' }^{(0)}    U_{   L_y,n }^{(0)\ast} \right)\left(   U_{0,n}^{(0)\ast}    U_{ L_y+1,n'}^{(0)}  -  U_{0,n'}^{(0)}    U_{ L_y+1,n}^{(0)\ast} \right) \right] ,  \\[8pt]
    \Gamma^{bc} 
      & =   -\frac{4}{ L_x} \sum_{0<q\leq \pi}
    \sum_{n,n'}' \frac{ 1}{\varepsilon^{(0)}_{n} - \varepsilon^{(0)}_{n'}}\left\vert   U_{0,n}^{(0)\ast}    U_{ L_y+1,n'}^{(0)}  -  U_{0,n'}^{(0)}    U_{ L_y+1,n}^{(0)\ast} \right\vert^2   ,  \\[8pt]
        \Gamma^{cb} 
      & =   -\frac{4}{ L_x} \sum_{0<q\leq \pi}
    \sum_{n,n'}' \frac{ 1 }{\varepsilon^{(0)}_{n} - \varepsilon^{(0)}_{n'}}\left\vert   U_{1,n}^{(0)\ast}    U_{ L_y,n'}^{(0)}  -  U_{1,n'}^{(0)}    U_{ L_y,n}^{(0)\ast} \right\vert^2   ,
\end{align} 
where the sum $\sum_{n,n'}^{\prime}$ runs for $1\leq n \leq L_y/2+1$ and $n'\neq n$. 

Equation (\ref{eq:SM-chi-Gamma}) becomes exact in the limit $\tilde{J} \to \tilde{J}_c^{+}$. A nontrivial solution with  $\chi_b \neq 0$ and $ \chi_c \neq 0$ requires
\begin{equation}
    \tilde{J}^2_{c}  \, \text{det }\Gamma   -   \tilde{J}_{c} \,  \text{tr }\Gamma  + 1   =   0   , 
\end{equation}
where $ \text{det }\Gamma   =    ( \Gamma^{bb} )^2 -  \Gamma^{bc} \Gamma^{cb}  $ and $\text{tr }\Gamma   =   2  \Gamma^{bb}    $.
From the general structure of the matrix elements of $\Gamma$ and using the Cauchy-Schwarz inequality, we can show that  $\text{det }\Gamma < 0$. We found numerically that  $\text{tr }\Gamma > 0$ for   $J>0$. Thus, the only positive solution is 
\begin{equation}
\tilde{J}_c   =    \frac{ \text{tr }\Gamma  }{ 2 \vert \det \Gamma \vert  }   \left[   -1   +   \sqrt{   1   +   \frac{4 \vert \det \Gamma \vert  }{\left(   \text{tr }\Gamma    \right)^2}  }  \, \right]           .
\end{equation}


\subsection{3.  Continuum limit of Hamiltonian and spin operators}

The uncoupled phase  is described by the mean-field Hamiltonian with $\chi_b=\chi_c=0$. Equivalently, we can set $\tilde{J}=0$ in Eq. (\ref{eq:SM-iA}). For $\kappa \neq 0$,   there are two gapless modes with linear dispersion about $q=0$ and all other modes are gapped.

To describe the low-energy physics,  we treat $q$ as a small parameter and expand the Hamiltonian matrix as
\be \im A(q) = \mathcal{H}^{(0)} + q \, \mathcal{H}^{(1)} + \dots. \ee
The matrix $\mathcal{H}^{(0)}$ is   tridiagonal   with coefficients depending on $h_z$ and $J$, whereas   $\mathcal{H}^{(1)}$ is a pentadiagonal matrix with a linear dependence on $\kappa$. 

We denote the two eigenvectors of $\im A(q)$ associated with the  gapless   modes   by $\phi_R(q)$ and $\phi_L(q)$. When we project  onto these low-energy modes,  Eq. (\ref{eq:SM-d-gamma}) reduces to 
\be d_{q,y}^{\phantom{\dagger}}  \sim     \phi_{R,y}(q)   \gamma_{q,R} + \phi_{L,y}(q)   \gamma_{q,L} \; \label{eq:SM-d-gamma-low} .\ee
We expand the eigenvectors   linearly in   momentum as
\be \phi_{R/L}(q) = \phi_{R/L}^{(0)} + q \; \phi_{R/L}^{(1)} + \dots . \ee
 For  $q\to0$, these vectors become the two eigenvectors of  $\mathcal{H}^{(0)}$ with zero eigenvalue. Consider $L_y$ even and   $L_y \gg 1$. For $\kappa > 0$, the non-zero components of these vectors are
\begin{align}
    \phi_{L,y}^{(0)} \; &= \;  \frac{h_z}{2J} \, (-1)^{\frac{y}{2}} \,  e^{-\frac{L_y-1-y}{\xi}}\, \phi_{L,L_y+1}^{(0)} \; ,  \quad  y  \text{ odd} ,\; y \leq L_y-1 \; , \label{eq:SM-phi_L} \\
    \phi_{R,y}^{(0)} \; &= \;  \frac{h_z}{2J} \, (-1)^{\frac{y}{2}} \,  e^{-\frac{y-2}{\xi}}\,\phi_{R,0}^{(0)} \; ,  \qquad  y  \text{ even}, \; y \geq 2 , \label{eq:SM-phi_R}
\end{align}
where the correlation length is $\xi = 2/\ln 2$ and the normalization factor is $\phi_{L,L_y+1}^{(0)} = \phi_{R,0}^{(0)} = -J \left( J^2  + h_z^2/3 \right)^{-1/2}$. The right- and left-moving modes are related by a  transformation that takes $y \mapsto L_y+1-y$ and $v\mapsto -v$.

The eigenvalue equation at first order in $q$ yields 
\be  \mathcal{H}^{(0)} \phi_{\alpha}^{(1)} +  \mathcal{H}^{(1)}  \phi_{\alpha}^{(0)}  = \alpha v \,  \phi_{\alpha}^{(0)}  \, , \label{eq:SM-1st} \ee 
where $\alpha = R/L = \pm$. 
Using that $\phi_\alpha^{(0)}$ has eigenvalue zero in $\mathcal{H}^{(0)}$, we obtain
\be v =[ \phi_R^{(0)}]^t \mathcal{H}^{(1)}  \phi_{R}^{(0)}   =  \frac{\vert \kappa \vert  h^2_z}{J^2 + h^2_z/3} \; .   \ee
Here the velocity is  positive by definition. For $\kappa > 0$, the right movers are localized at the $\ell_1$ edge  ($y=1$) and the left movers at the the $\ell_2$ edge  ($y=L_y$). If we vary the magnetic field so that $\kappa$ changes sign,  the Chern number in the bulk also changes sign and the chirality at the edge is reversed, and we must relabel $R \leftrightarrow L$ in Eqs.  (\ref{eq:SM-phi_L}) and (\ref{eq:SM-phi_R}).  

For $\kappa>0$, the spin operator at    edges $\ell_1$ and $\ell_2$ are
\begin{align}
    S^z_{R}(x) &= \frac{\im}{2} d_{x,0}d_{x,1} = \frac{\im}{L_x} \sum_{-\pi< q, k \leq \pi } e^{\im (q+k) x} \, d_{q,0} d_{k,1} \\
    S^z_{L}(x) &= \frac{\im}{2} d_{x,L_y+1}d_{x,L_y} = \frac{\im}{L_x} \sum_{-\pi< q, k \leq \pi } e^{\im (q+k) x} \, d_{q,L_y+1} d_{k,L_y}   . 
\end{align}
Substituting the projected mode expansion in Eq. (\ref{eq:SM-d-gamma-low}), we find that the term of zeroth order in $q$ vanishes, and we need to consider the first-order corrections. As the first and last rows of  $\mathcal{H}^{(1)}$ are zero,  the first and last components of Eq. (\ref{eq:SM-1st}) yield a simple relation between the required components of $\phi_{\alpha}^{(0)}$ and $\phi_{\alpha}^{(1)}$. We find   
\begin{align}
    \phi_{R,1}^{(1)} & = \frac{v}{-\im h_z} \, \phi_{R,0}^{(0)} = \frac{Jv}{\im h_z}  \left( J^2  + h_z^2/3 \right)^{-1/2} \; ,
    \\[5pt]
    \phi_{L,L_y}^{(1)} & = \frac{-v}{-\im h_z} \, \phi_{L,L_y+1}^{(0)} = -\frac{Jv}{\im h_z}  \left( J^2  + h_z^2/3 \right)^{-1/2} \; .
\end{align}
Using Eq. (\ref{eq:SM-d-gamma-low}) and fact that the low-energy modes are localized at the edge, we can write the spin operators   as 
\begin{align}
    S^z_{R}(x)&=\frac{\im}{L_x}\sum_{q, k}e^{\im (q+k) x}\;\phi_{R,0}(q) \phi_{R,1}(q) \gamma_{q,R} \gamma_{k,R} \; ,\\
    S^z_{L}(x)&=\frac{\im}{L_x}\sum_{ q, k  }e^{\im (q+k) x}\;\phi_{L,L_y+1}(q) \phi_{R,L_y}(q) \gamma_{q,L} \gamma_{k,L}\; .
\end{align}
The coefficients up to linear order in $q$ are 
\begin{equation}
    \phi_{R,0}(q) \phi_{R,1}(q) =  - \phi_{L,L_y+1}(q) \phi_{R,L_y}(q) = q \;\frac{\im J^2 v}{h_z} \left( J^2  + \frac{h_z^2}{3} \right)^{-1} \equiv isq.
\end{equation}
Taking the Fourier transform back to real space in the continuum, we can write $S^z_\alpha(x)$ in terms of the chiral Majorana fermions $\gamma_\alpha(x)$ as  in Eq. (10) of the main text.


\subsection{4. Calculation of dynamic spin correlations at finite temperature}

Consider the    spin-spin correlation in imaginary time:
\begin{equation}
        \tilde{\chi}_{\alpha}(x-x',\tau)   =  - \left\langle \, T_\tau S_{\alpha}^{z}(x,\tau) S_{\alpha}^{z}(x',0) \,  \right\rangle_{\beta}  =   \frac{s^2}{4}   \Big\langle T_\tau   \gamma_{\alpha}(x,\tau) \partial_x \gamma_{\alpha}(x,\tau)  \gamma_{\alpha}(x',0) \partial_{x^{\prime} } \gamma_{\alpha}(x',0)    \Big\rangle_{\beta}    , 
\end{equation}
where $\beta$ in the inverse temperature and $T_\tau $ denotes time ordering. Its Fourier transform is the dynamical spin susceptibility $\chi(q, \im \omega_l)$, where $\omega_l$ are bosonic Matsubara frequencies.  In the calculation of the spin-lattice relaxation rate, we need the local spin susceptibility, which may be obtained as
\begin{align}
    \sum_{q}  \chi_{\alpha}(q, \im \omega_l )  =  \frac{L_x}{2} \int_{-\beta}^{\beta} d\tau  \, e^{\im \omega_l \tau } \,  \tilde{\chi}_{\alpha}(0, \tau ).
\end{align}
Using Wick's theorem, we can express the spin-spin correlation in terms of the  Green's function  $\mc G_\alpha(x,\tau)=-\langle T_\tau  \gamma_\alpha(x,\tau)\gamma_\alpha(0,0) \rangle$  for   noninteracting Majorana fermions. In frequency-momentum space,  we have $
        \mathcal{G}_{\alpha}(q , \dot{\imath} \omega_{n} )    =     (\dot{\imath} \omega_{n} - v_{\alpha}q)^{-1}   $, where $v_\alpha=\alpha v$.  
Performing the sum over internal Matsubara frequencies, we obtain    
\begin{equation}
    \begin{split}
     \sum_{q} \chi_{\alpha}(q, \dot{\imath} \omega_l )    =    \frac{s^2}{4  L_x}    \sum_{k_{1},k_{2} \geq 0} &     \Bigg\{      k_{2}(k_{2}-k_{1})      \frac{ f(  v_{ \alpha } k_{1} ) - f( - v_{ \alpha } k_{2} )  }{\dot{\imath} \omega_{ \ell  }  -v_{ \alpha } k_{1} - v_{ \alpha } k_{2} } 
        +   k_{2} (k_{1}+k_{2})      \frac{ f(  v_{ \alpha } k_{1} ) - f(  v_{ \alpha } k_{2} )  }{\dot{\imath} \omega_{ \ell  } - v_{ \alpha } k_{1}+  v_{ \alpha } k_{2}}          \\[6pt]
        &  -      k_{2}(k_{2}+k_{1})       \frac{ f( v_{ \alpha } k_{1}) - f(  v_{ \alpha } k_{2} )}{\dot{\imath} \omega_{ \ell  }  + v_{ \alpha } k_{1}- v_{ \alpha } k_{2}}   
        -      k_{2}(k_{2}-k_{1})   \frac{ f( v_{ \alpha } k_{1} ) - f( -  v_{ \alpha } k_{2} )  }{\dot{\imath} \omega_{ \ell  } + v_{ \alpha } k_{1}+  v_{ \alpha } k_{2} }     \Bigg\}   , 
    \end{split} 
\end{equation}
where $   f(\omega)   =   \left( e^{\beta \omega} +1 \right)^{-1}$ is the Fermi-Dirac distribution. Taking the analytic continuation to real frequencies, $\dot{\imath} \omega_{l} \to  \omega + \dot{\imath} 0^{+}$ and the limit $ L_x \to \infty$, we write the imaginary part of the retarded spin susceptibility  as\begin{equation}
     -2\sum_{q}  \mathrm{Im}    \chi_{\alpha}^{\text{ret}}(q, \omega)    =   - \frac{  s^2}{4 v^2} \left[ \mathcal{I}(\omega,T) - \mathcal{I}(-\omega,T) \right],
\end{equation}
where we define the integral
\begin{equation}
    \mathcal{I}(\omega,T) =  \int_{-\infty}^{\infty} dk  \,    k( \omega +2 v_{\alpha} k )      \Big[ f( \omega + v_{ \alpha } k ) - f(  v_{ \alpha } k ) \Big]   . 
\end{equation}
At zero temperature,   the Fermi-Dirac distribution reduces to a step function and we obtain $\mathcal{I}(\omega,0) = - \omega^3/6v^2$. Using the Sommerfeld expansion for a quadratic function,
\begin{align}
    \int_{-\infty}^{\infty} d\varepsilon   \left( A_2 \varepsilon^2 + A_1 \varepsilon + A_0 \right)  \left[  
    f(\varepsilon- \mu) -  \Theta(-\varepsilon)   \right]   &=   
    A_2 \left( \frac{\mu^3}{3}  +  \mu \pi^2 \frac{T^2}{3} \right) 
    + A_1 \left(  \frac{\mu^2}{2}  +  \pi^2 \frac{T^2}{6} \right)
    + A_0    \mu   ,
\end{align}
we can show that 
\begin{equation} 
     \mathcal{I}(\omega,T)   =- \frac{\omega^3}{6v^2} - \frac{2 \pi^2}{3v^2}   \omega   T^2   .
\end{equation}
Therefore, the dynamical spin structure factor at low temperatures  is given by  \begin{equation}
     S(q,\omega_0) \approx -\frac{2T}{\omega_0}\sum_{q}  \mathrm{Im}    \chi_{\alpha}^{\text{ret}}(q, \omega_0)  = \frac{  s^2}{12 v^4}  \Big(  \omega^2_0 T   +   4 \pi^2 T^3    \Big)   .
\end{equation}

\end{document}